\documentclass[12pt,a4paper,final]{iopart}

\usepackage{iopams}  
\usepackage{graphicx}
\usepackage[breaklinks=true,colorlinks=true,linkcolor=blue,urlcolor=blue,citecolor=blue]{hyperref}
\usepackage[subrefformat=parens,labelformat=parens]{subfig}

\begin{document}

\title{Branch crossing behavior and interface states in 1D Inversion Symmetric Quaternary Photonic Crystals using Effective Medium Theory}

\author{Nicholas J. Bianchi$^{1}$}
\address{$^1$University of Rhode Island, Kingston Rhode Island}
\ead{nicholasbianchi@uri.edu}

\author{Leonard M. Kahn$^{1}$}
\address{$^1$University of Rhode Island, Kingston Rhode Island}
\ead{lenkahn@uri.edu}

\begin{abstract}
Effective medium theory is used to model a one dimensional lossy dielectric quaternary photonic crystal as a homogeneous slab. The unit cell of the original crystal is inversion symmetric  with layer sequence ACBCA. The behavior of the branch frequency singularities in the effective refractive index is investigated as  parameters in the layered structure change. A heterostructure composed of multiple photonic crystals is also modeled with effective medium theory. It is shown that the  effective refractive index of such a structure possesses regions of normal dispersion that correspond to localized interface states within the overall regions of anomalous dispersion representing the photonic band gap.
\end{abstract}

\vspace{2pc}
\maketitle

\section{Introduction}

Effective medium theory (EMT) is the  modeling of scattering parameters, such as reflection and transmission, from a complex homogeneous structure by replacing it with a partially or fully homogenized material, in an effort to simplify a problem. While any field of wave mechanics can make use of EMT, it has seen significant applications in electromagnetism (EM).  In 1956, Rytov \cite{Rytov} proposed that a one dimensional Bragg grating composed of two different isotropic layers could be thought of as a uniform slab with anisotropic effective permittivity and permeability. Since then,  EMT has been used to model both photonic crystals \cite{Yablonovitch,John}, which have scattering elements on the order of the incident wavelength, and metamaterials \cite{Smith3}, where wavelength is several orders of magnitude larger. In the long wavelength regime, EMT was applied to 2D  \cite{Lalanne, Kikuta, Chern} and 3D \cite{Ono,Datta} periodic photonic crystals (PCs) by approximating them as 1D Bragg gratings. Reflection from a 1D Bragg grating at oblique angles using EMT was examined \cite{Tang, Kameda}.

EMT has also been applied the photonic systems for higher frequencies. In 1976, Yariv and Yeh \cite{Yariv2} used an effective index profile representation of a binary PC whose layers have the same optical path to achieve nonlinear phase matching. It was shown that when a PC is homogenized, the resultant slab behaves as a single negative material inside the photonic band gaps (PBGs) \cite{Yong}.  Later, it was found that a bi-anisotropic parameter (magnetoelectric coupling) is present  when homogenizing unit cells without inversion symmetry \cite{Liu1,Liu2}. These works also showed that power expansions in frequency for effective parameters becomes invalid at the start of the first PBG. This breakdown was due to the presence of branch points (i.e. singularites) \cite{Liu1}. Liu \cite{Liu1} also showed that effective index and impedance can be modeled with the Kramers-Kroing relations. Various concepts, such as density of states \cite{Boedecker}, reflection phase \cite{Zhao,Gao}, and defects \cite{Pei} have been studied in PCs using EMT.


In this work, we base the design of the laminated PCs from Ref.~\cite{Bianchi}. The individual layers in the symmetric unit cell are denoted as $A, B,$ and $C$ with refractive indices $n_A, n_B,$ and $n_C$. The permittivity, $\epsilon$ and permeability, $\mu$ as defined as $n_i^2 = \epsilon_i \mu_i$  The period, $\Lambda$ and optical path, $\Gamma$, across a unit cell are given as constants, where $\gamma = \Gamma/\Lambda$. As a result, the scaled layer lengths are,

\begin{equation}
d_A = \textrm{Re}\frac{\gamma - n_B - 2(n_C - n_B)d_C}{n_A - n_B}   \label{dA}
\end{equation}

\begin{equation}
d_B =  \textrm{Re}\frac{\gamma - n_A - 2(n_C - n_A)d_C}{n_B - n_A}    \label{dB}
\end{equation}

From the given  $\epsilon$ and $\mu$ in the PC, the transfer matrix method \cite{Soukoulis,Yariv1} is used to calculate transmission and reflection spectra. Then, applying EMT, these spectral functions are used to recover effective index and impedance relations in the associated uniform slab.
The primary purpose of this work is to examine the behavior of branch intersections in the effective refractive index as parameters in the corresponding laminated PC vary. It is a well known problem that recovering an effective refraction index from transmssion and reflection functions introduces extraneous solutions in the form of multiple branches since the cosine function is multi-valued \cite{Smith1}. The goals of this work are to: (1) review the methods of effective medium theory for one-dimensional systems and compare the analytic and numerical techniques for finding effective parameters, (2) apply EMT to quaternary PCs with tunable parameters to show how branch intersections evolve as said parameters change, and (3) describe a heterostructure composed of binary and quaternary PCs in terms of EMT and explain interface modes in such a system in terms of the effective parameters.

\begin{figure}[!ht]
	\center
	\subfloat[]{\includegraphics[width=0.6\columnwidth]{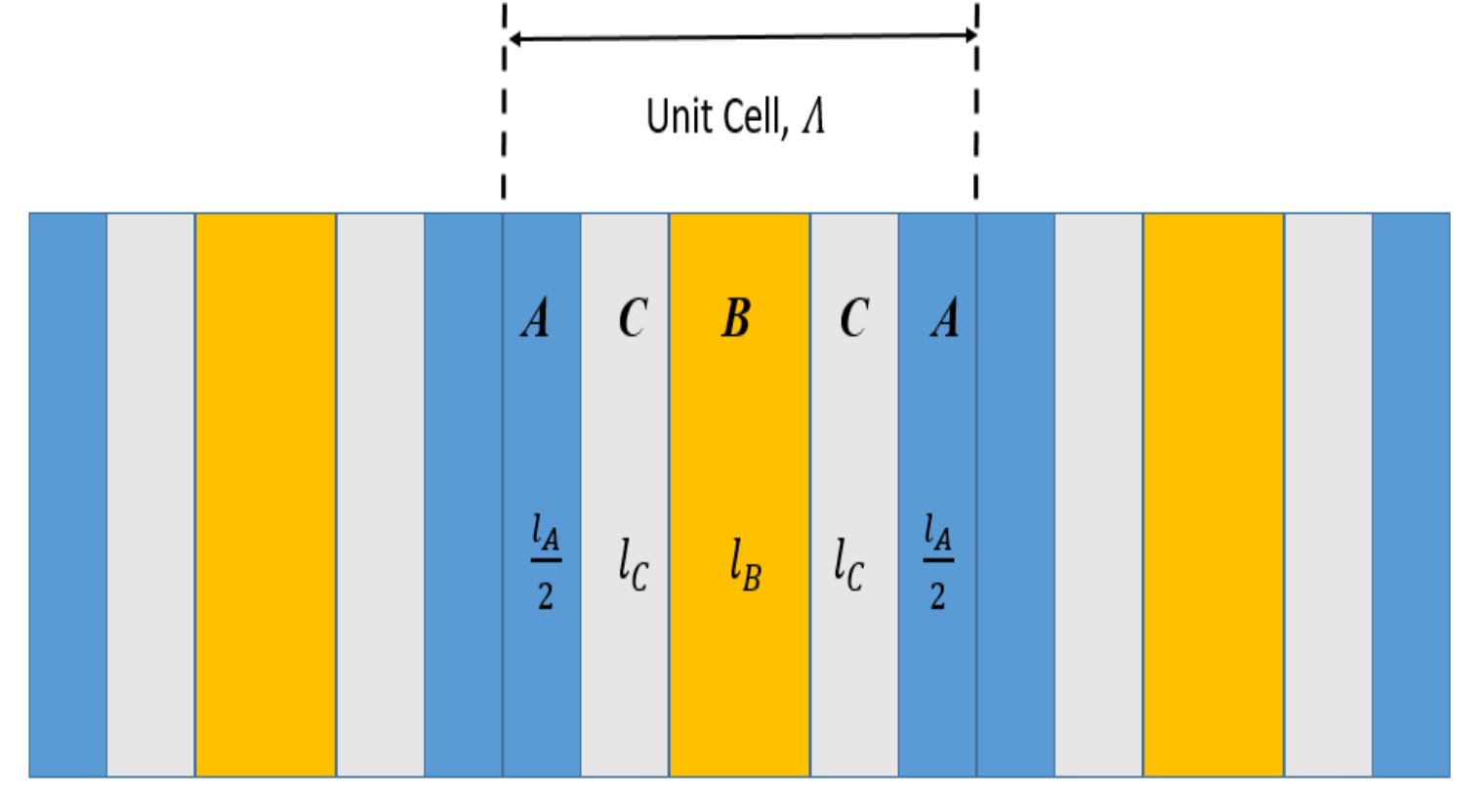}   \label{PC_diagram} }
	\caption{Inversion symmetric quaternary photonic crystal }
	\label{QPC}
\end{figure}

\section{Methods}

While in most scattering problems the objective is to find reflection and transmission spectra based on the given material parameters (\textit{i.e.} $\epsilon ~\&~ \mu$ or $n ~\& ~z$), with effective medium theory the inverse problem can be performed. In this case, a system has a known or measured spectrum. With this information, a more uniform object can be considered that produces the same scattering. Essentially, this means that for the uniform material to return the same scattering information as the heterogeneous one, the object must act as though it has frequency dependent dispersion rather than spatial dispersion.

Smith \textit{et. al}\cite{Smith1}  provided an analytical method for extracting the effective refractive index and impedance from the transmission and reflection coefficents, $t$ and $r$ respectively. Matching electric and magnetic boundary conditions for a single slab yields,

\begin{equation}
t^{-1} = \cos\phi - \frac{i}{2}\left(z + \frac{1}{z} \right)\sin\phi  \label{t}
\end{equation}

\begin{equation}
r t^{-1} = -\frac{i}{2}\left(z - \frac{1}{z} \right)\sin\phi             \label{r}
\end{equation}
where $\phi = n_{eff} k_0 L$ for freespace wavevector, $k_0 = \omega/c$, and slab length, $L$. Using dimensionless variables, $\phi = 2\pi n_{eff} \xi N$. The length must be an integer multiple, $N$, of the unit cell period, $\Lambda$, and $\xi = f\Lambda/c$ is the scaled frequency. Equations \ref{t} and \ref{r} can then be inverted to produce,

\begin{equation}
z^2 = \frac{(r+1)^2-t^2}{(r-1)^2-t^2} \label{z_eff}
\end{equation}

\begin{equation}
\cos\phi = X(\xi) =  \frac{1}{2t}(1+t^2-r^2) \label{n_eff}
\end{equation}
Finding the permittivity and permeability is now trivial, $\epsilon = n/z$ and $\mu = nz$.

It is assumed that the dielectrics in the original unit cell can only have lossy components. Therefore, the positive root is chosen for the impedance in Eq.~\ref{z_eff}, since the materials are passive \cite{Smith1}. While symmetric unit cells produce well defined, unique values of impedance, it was found that when homogenizing an asymmetric cell, impedance will depend on the order of the layers. This discrepancy is due to the diagonal scattering matrix elements, $S_{11}$ and $S_{22}$ having different amounts of phase advance \cite{Smith2}. Calculating the effective refractive index is more difficult due to the presence of the cosine function. Mathematically, there are infinite values of $n_{eff}$ that satisfy Eq.~\ref{n_eff}, but, physically, there should only be one.  As before, to avoid phase ambiguity, it is best to consider symmetric unit cells, although $n_{eff}$ dependence on unit cell asymmetry is less pronounced compared with $z_{eff}$ \cite{Smith2}. Since the original system is periodic, the thickness of the slab should be taken as small as possible to minimize the number of roots in $\cos\phi$; however, since $L = N \Lambda$, the smallest value of $L$ is a single unit cell ($N=1$) \cite{Smith1, Smith2, Chen2}.

\begin{figure}[!ht]
	\subfloat[][]{\includegraphics[width=0.5\columnwidth]{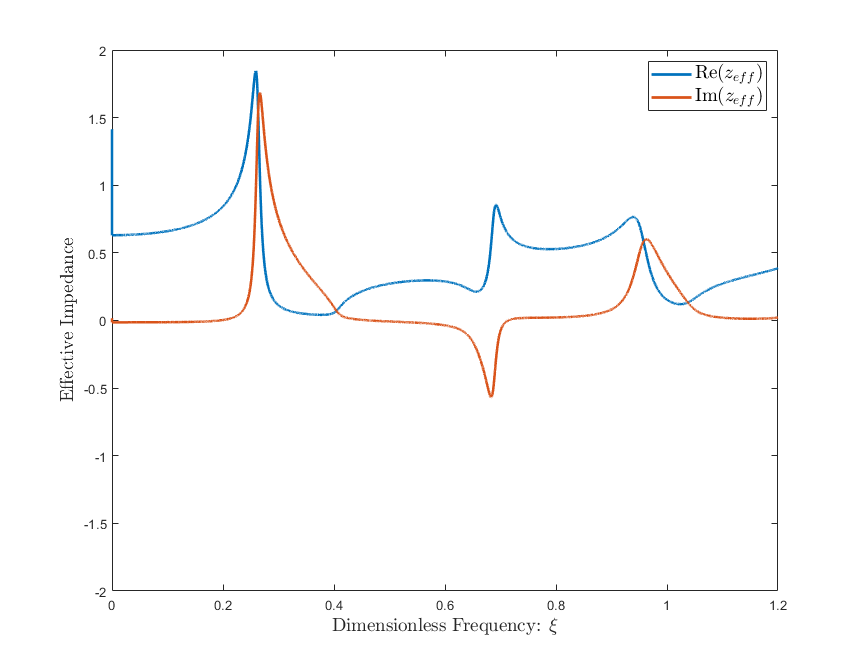}   \label{Eff_Z} }
	\subfloat[][]{\includegraphics[width=0.5\columnwidth]{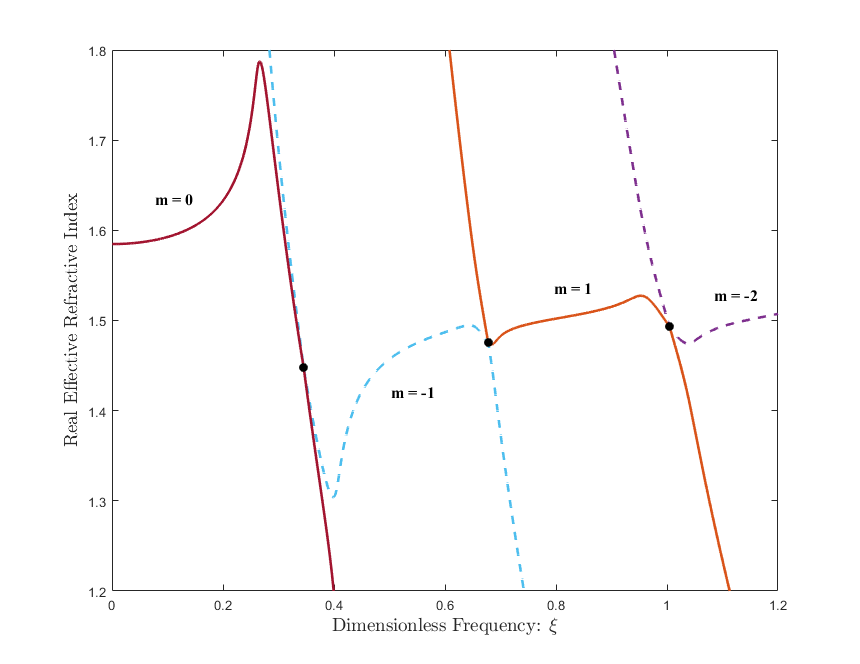}   \label{Real_n_1} }\\
	\subfloat[][]{\includegraphics[width=0.5\columnwidth]{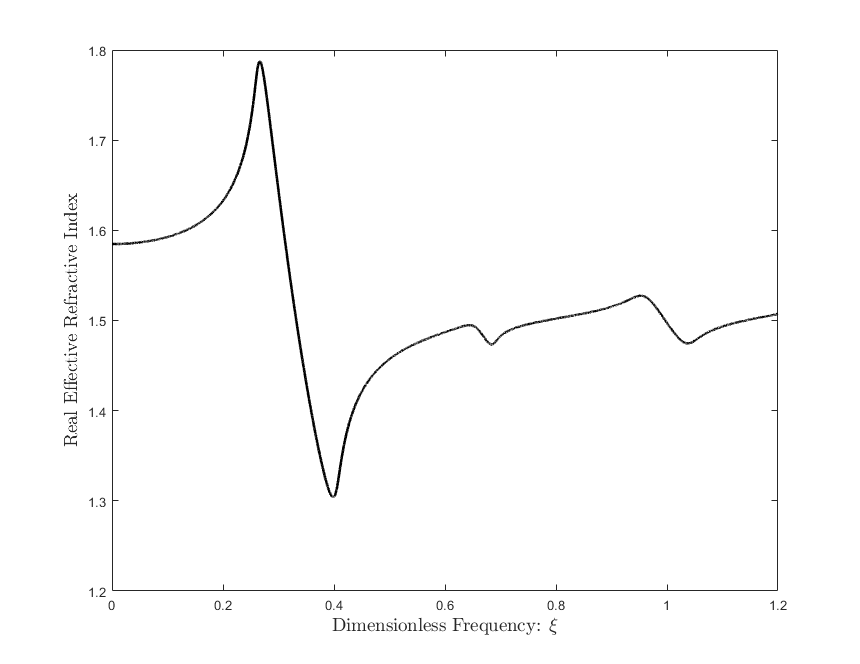}   \label{Real_n} }
	\subfloat[][]{\includegraphics[width=0.5\columnwidth]{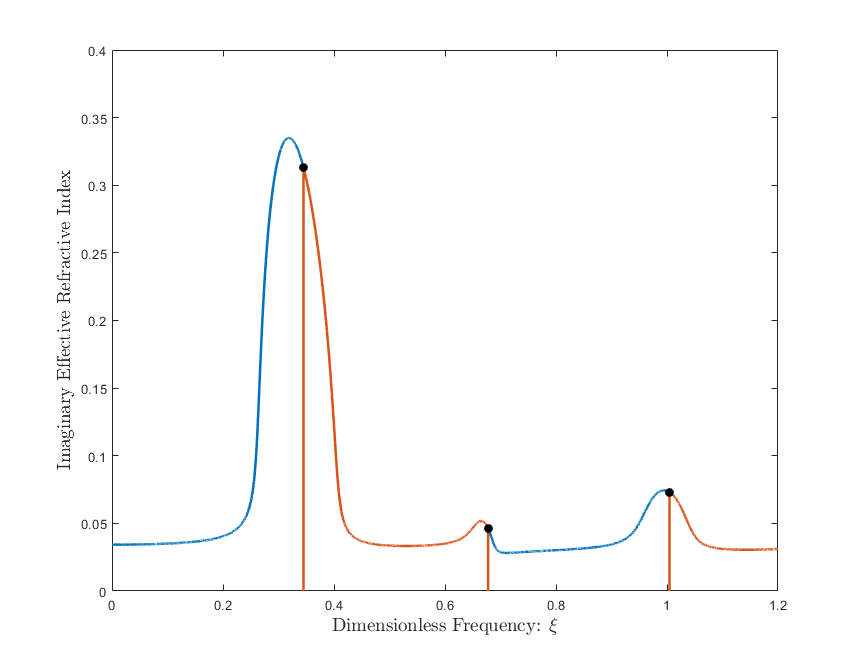}   \label{Imag_n_1} }
	\caption{Example of effective parameters of 1 unit cell from Fig.~\ref{PC_diagram}. In the original unit cell $d_A=0.14207$, $d_B=0.45793$, $d_C=0.2$, $\epsilon_A=6+0.2i$, $\mu_A=1$, $\epsilon_B=1$, $\mu_B=1$, $\epsilon_C=3+0.2i$, and $\mu_C=1$.(a) Real and imaginary effective impedance. (b) Real component of refractive index. The values of $m$ specify the branches to ensure $n_R$ is continuous and smooth. The black dots highlight the points where the branches connect. (c) Real component of refractive index after nonphysical solutions are removed. (d)  Imaginary component of refractive index. }
	\label{Analytical_n}
\end{figure}

If we let $n_{eff} = n_R + i n_I$ and $X = X_R + i X_I$, Eq.~ \ref{n_eff} can be separated into \cite{Smith1},

\begin{equation}
n_R(\xi) = \frac{1}{2\pi N \xi} \left[ \textrm{Re}(\cos^{-1}( X(\xi) ) ) +2 \pi m \right]  \label{n_eff_real}
\end{equation}
and
\begin{equation}
n_I(\xi) = \frac{1}{2\pi N \xi}  \textrm{Im}(\cos^{-1}( X(\xi) ) )   \label{n_eff_imag}
\end{equation}
where $m$ is an integer that denotes the current branch. Figure \ref{Analytical_n} shows an example of the effective parameters for a photonic crystal with inversion-symmetric quaternary unit cells. Only 1 unit cell is used. Figure \ref{Eff_Z} shows the effective impedance using Eq.~\ref{z_eff}. In Fig.~\ref{Real_n_1}, which displays Eq.~\ref{n_eff_real}, branch crossing points, denoted by black dots, only occur in regions of anomalous dispersion, where $\frac{dn_I}{d\xi}<0$. Figure \ref{Real_n} displays the real part but without the extra branches. Some early attempts at plotting Re$(n_{eff})$ have resulted in erroneous solutions, such as incorrect branch transitions \cite{Notomi}, and incorrect or missing anomalous behavior \cite{Dowling,Jeong}. The corresponding crossing points are shown in the imaginary part in Fig.~\ref{Imag_n_1}. The positions of these points can be found by setting $X_I(\xi) = 0$ and are independent of $N$. This equation will be used in the next section to explore a rich supply of behaviors of the branch crossing points.  Increasing $N$, however, would cause additional crossing points to appear in the regions of normal dispersion, where $\frac{dn_R}{d\xi}>0$, with the number of crossing points in these region equal to $N-1$. 

\begin{figure}[!ht]
	\subfloat[][]{\includegraphics[width=0.5\columnwidth]{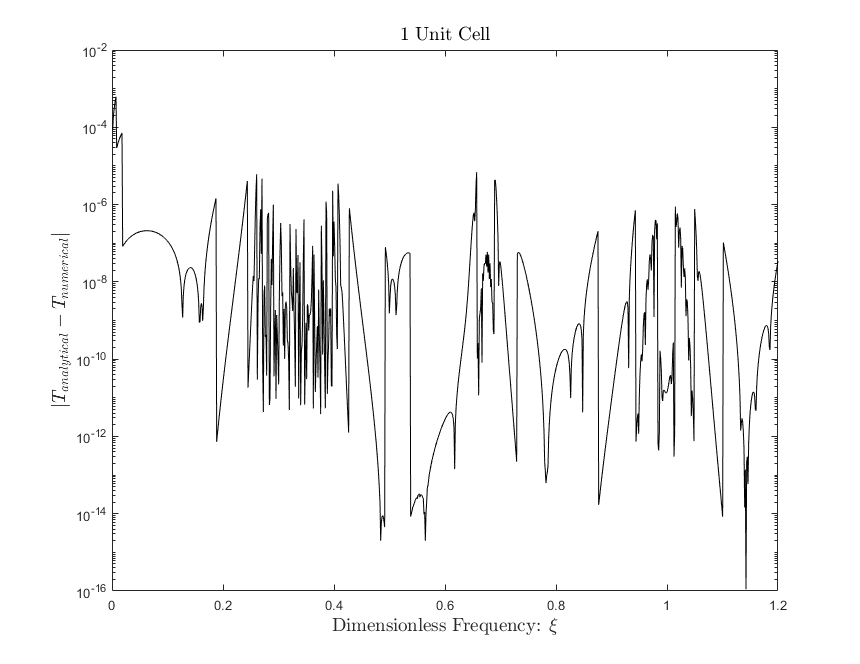}   \label{Trans_1} }
	\subfloat[][]{\includegraphics[width=0.5\columnwidth]{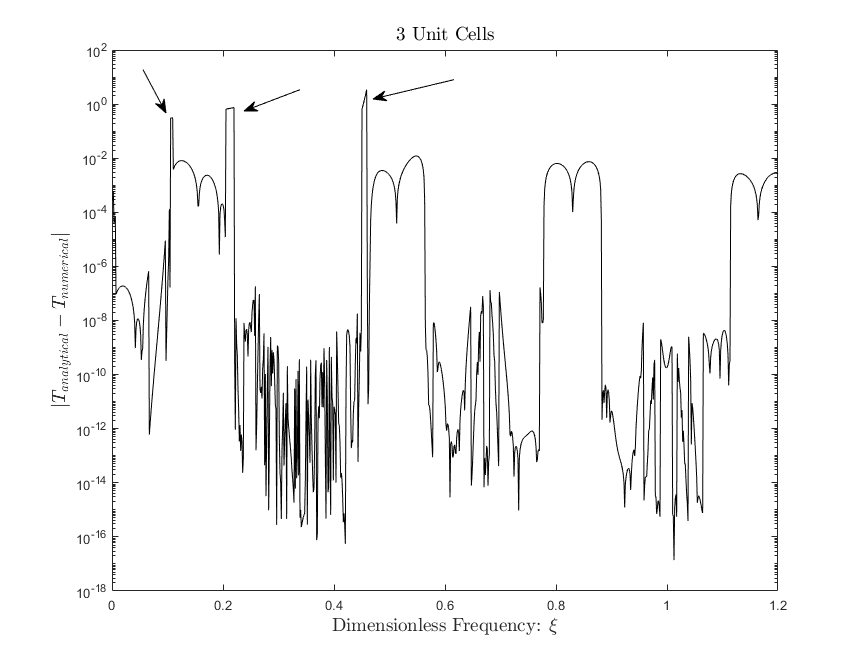}   \label{Trans_3} }
	\caption{Difference in transmission calculated analytically with Eqs.~\ref{n_eff_real} and \ref{n_eff_imag} and numerically with Eqs.~\ref{n_real_num} and \ref{n_imag_num}  (a) 1 unit cell  (b) 3 unit cells. Note the regions given by the arrows where convergence is poor. With 3 unit cells the band gaps are more visible, showing better agreement than surrounding areas.}
	\label{Trans_Diff}
\end{figure}

An alternative method to find $n_R$ and $n_I$ is to calculate them numerically. Starting from Eq.~\ref{n_eff}, we can separate out the real and imaginary equations \cite{Soukoulis},
\begin{equation}
\cos(2 \pi N \xi n_R)\cosh(2 \pi N \xi n_I) = X_R(\xi)       \label{n_real_num}
\end{equation} 
\begin{equation}
\sin(2 \pi N \xi n_R)\sinh(2 \pi N \xi n_I) = -X_I(\xi)      \label{n_imag_num}
\end{equation} 
Note that these equations are both coupled and nonlinear, making solutions challenging to find. As before, the imaginary component is easier to obtain since it is only located in hyperbolic functions, meaning that there is only a finite number of possible values it can have; however, we see again that the real component only appears in circular trigonometric functions, meaning we must be careful to select the correct branch. To proceed, it is best to start at zero frequency or at least very close to it. Here all non-physical solutions will diverge. The frequency $\xi$ is run through a list of values and both components are calculated at each step, gradually building an array of values. The main difficulty is selecting an appropriate initial starting value each interation so that $(n_R,n_I)$ converges to the correct solution. While the numerical method  can still be used at higher frequencies (or higher $N$), its practicality starts to falter as possible solutions become closer and closer together, making selecting the correct one more of a challenge. Figure \ref{Trans_Diff} illustrates this problem by logarithmically plotting the difference between the analytical and numerical transmission spectra for 1 and 3 uinits cells. Crystal parameters are the same as in Fig.~\ref{Analytical_n}. In Fig.~\ref{Trans_1}, a single unit cell is used, allowing the arguments of the sine and cosine funtions in Eqs.~\ref{n_real_num} and ~\ref{n_imag_num} to remain small enough to allow for proper convergence of $n_R$ and $n_I$ for all frequencies. In Fig.~\ref{Trans_3}, three unit cells are used, and it can be clearly seen that there are three regions, marked by arrows, in which agreement is poor. This is a result of Eqs.~\ref{n_real_num} and ~\ref{n_imag_num} converging to an incorrect solution for those frequency ranges. To remedy this problem for higher $\xi$ (or $N$), the same initial $(n_R,n_I)$ should not be used throughout the simulation. If it appears that $(n_R,n_I)$ will converge to non-physical values for a certain frequency range, then the initial point can be altered for that range.

\section{Results and Discussions}

The question now is, how do the crossing frequencies in Fig.~\ref{Real_n_1} evolve as different parameters in the quaternary PC change? As mentioned in the previous section, constructing the real part of the effective refractive index requires knowing the appropriate branches such that the index is both continuous and smooth. By understanding how properties such as material loss can  influence the location of these crossings, the construction of effective index functions can be improved, while reducing the possibility of connecting adjacent branches incorrectly.

\begin{figure}[!ht]
	\subfloat[][]{\includegraphics[width=0.5\columnwidth]{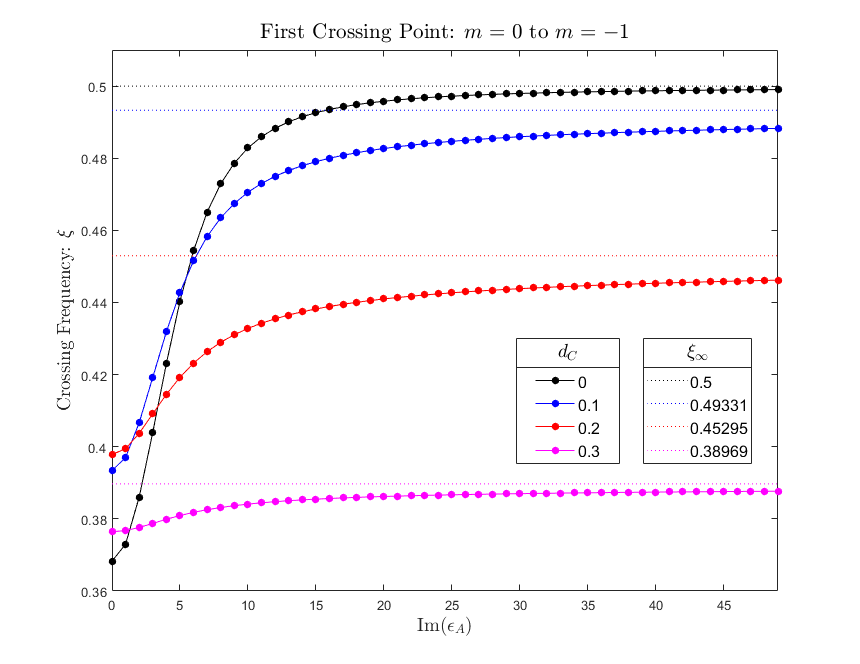}   \label{3_Im_Freq_1} }
	\subfloat[][]{\includegraphics[width=0.5\columnwidth]{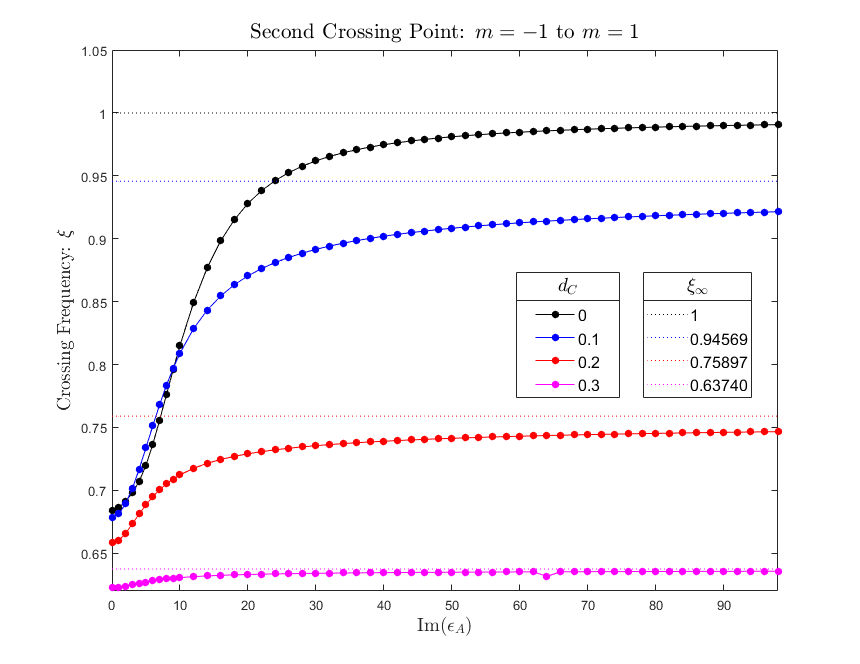}   \label{3_Im_Freq_2} }\\
	\subfloat[][]{\includegraphics[width=0.5\columnwidth]{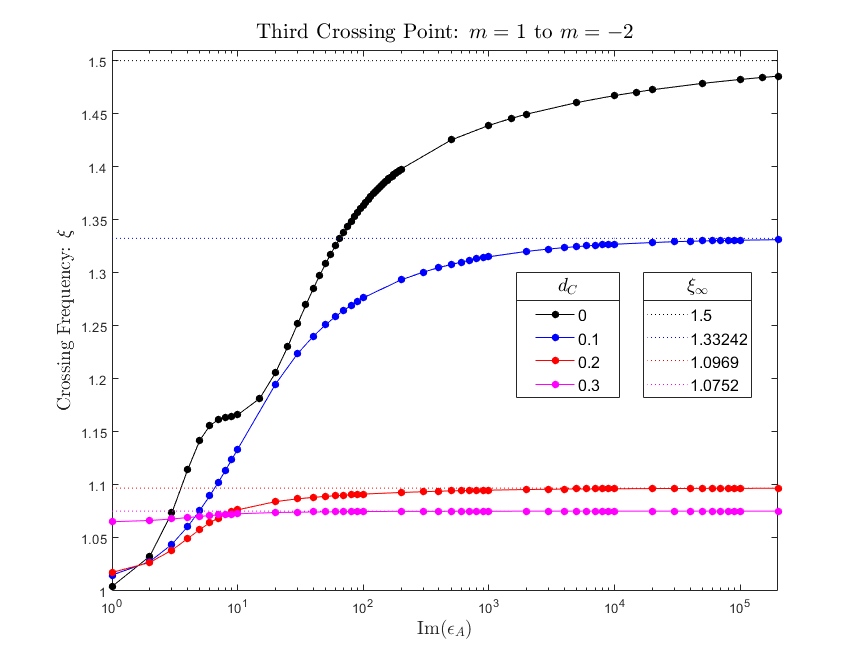}   \label{3_Im_Freq_3} }
	\caption{Behavior of (a) $1^{st}$  (b)$2^{nd}$  (c) $3^{rd}$ crossing point as \textrm{Im}($\epsilon_A$) changes}
	\label{3_ImA_Freq}
\end{figure}

\begin{figure}[!ht]
	\subfloat[][]{\includegraphics[width=0.5\columnwidth]{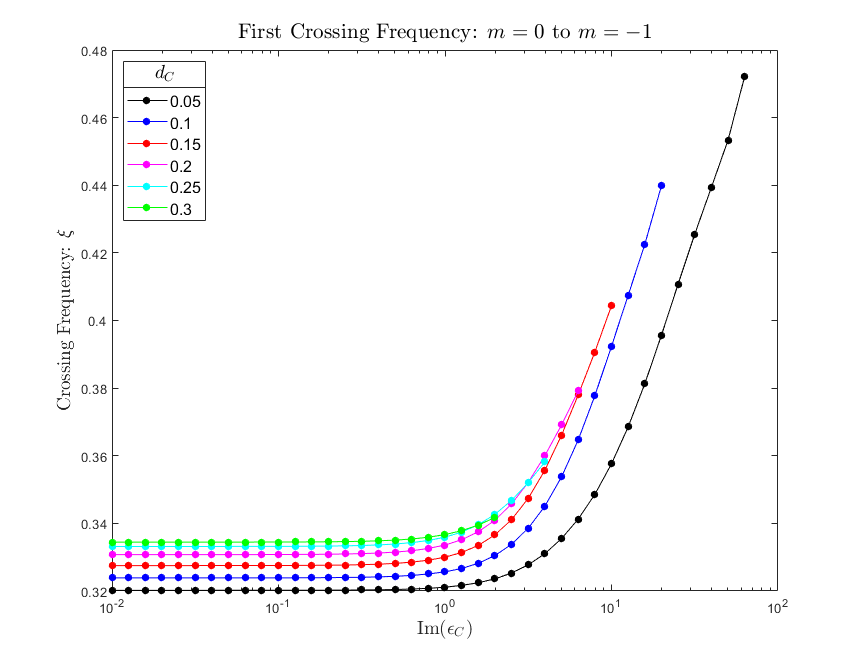}   \label{3_First_sing_ImC} }
	\subfloat[][]{\includegraphics[width=0.5\columnwidth]{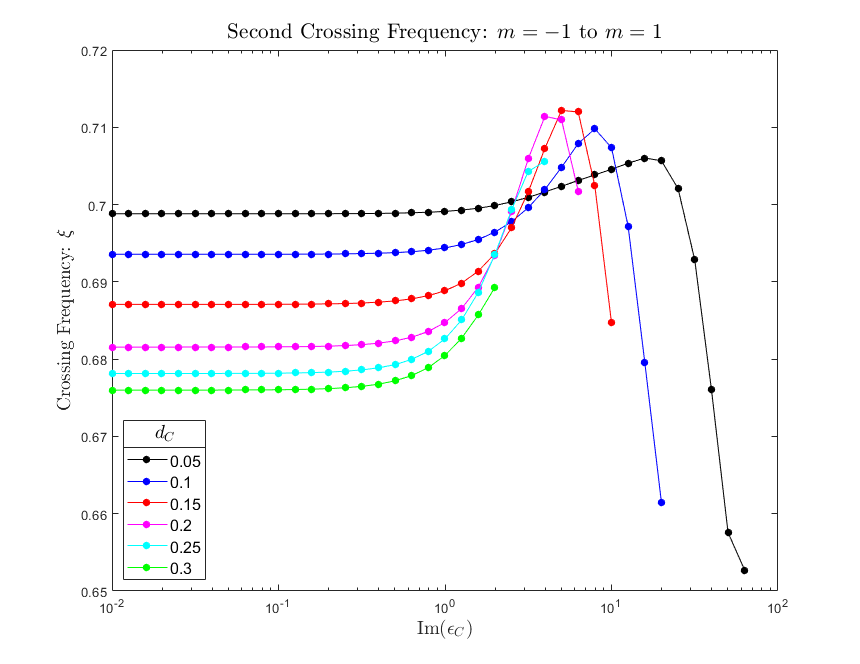}   \label{3_Second_sing_ImC} }\\
	\subfloat[][]{\includegraphics[width=0.5\columnwidth]{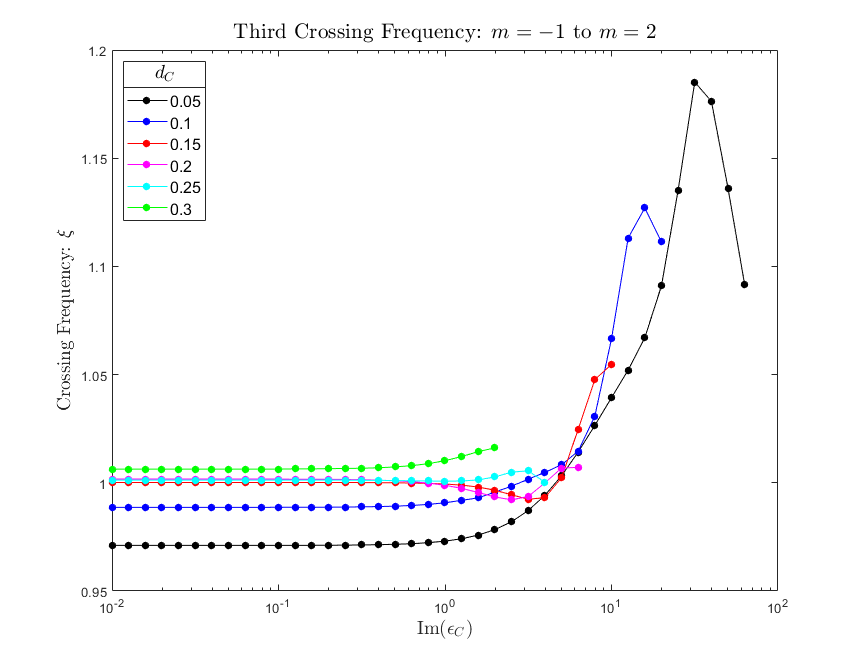}   \label{3_Third_sing_ImC} }
	\caption{Behavior of (a) $1^{st}$  (b)$2^{nd}$  (c) $3^{rd}$ crossing point as \textrm{Im}($\epsilon_C$) changes}
	\label{3_ImC_Freq}
\end{figure}

\begin{figure}[!ht]
	\subfloat[][]{\includegraphics[width=0.5\columnwidth]{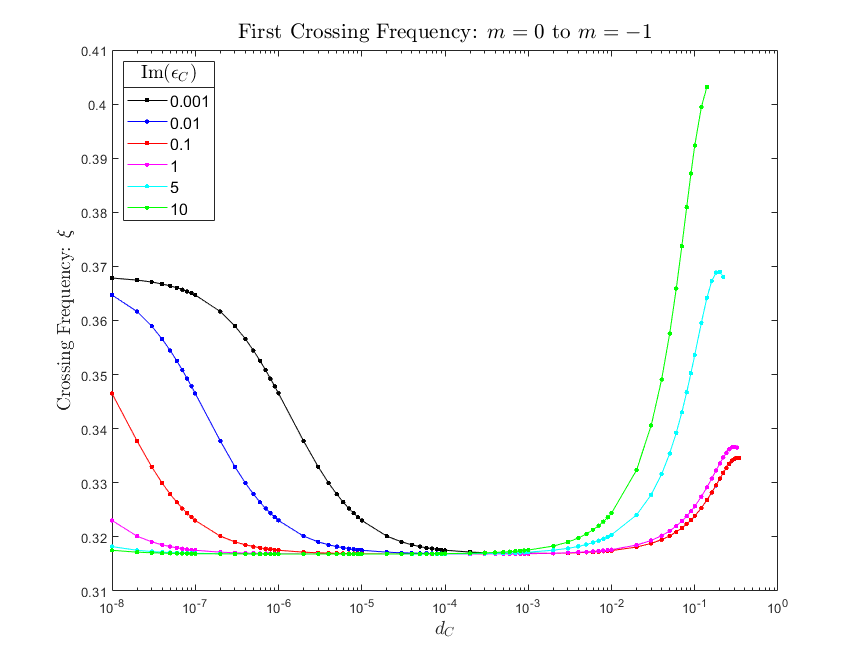}   \label{3_First_sing_dC} }
	\subfloat[][]{\includegraphics[width=0.5\columnwidth]{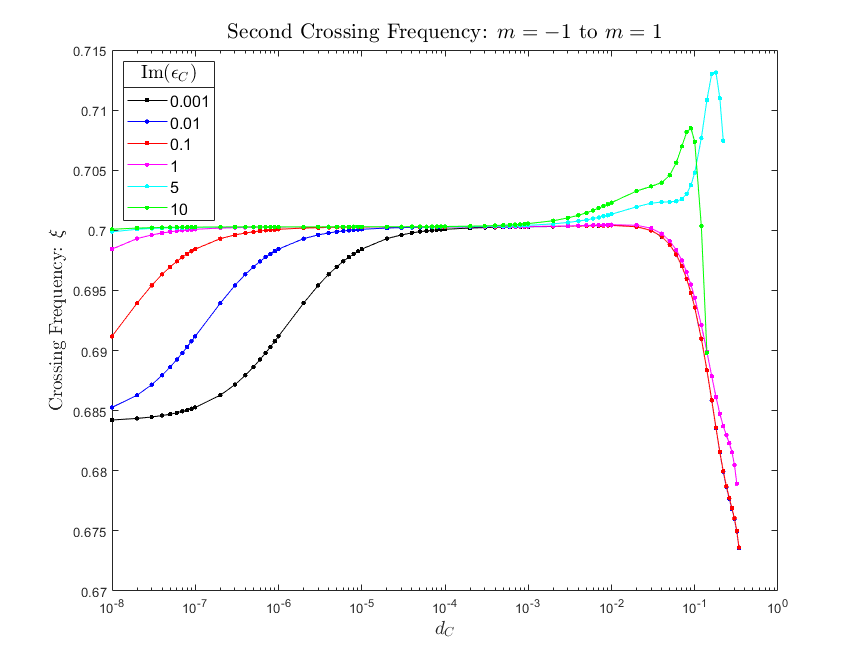}   \label{3_Second_sing_dC} }\\
	\subfloat[][]{\includegraphics[width=0.5\columnwidth]{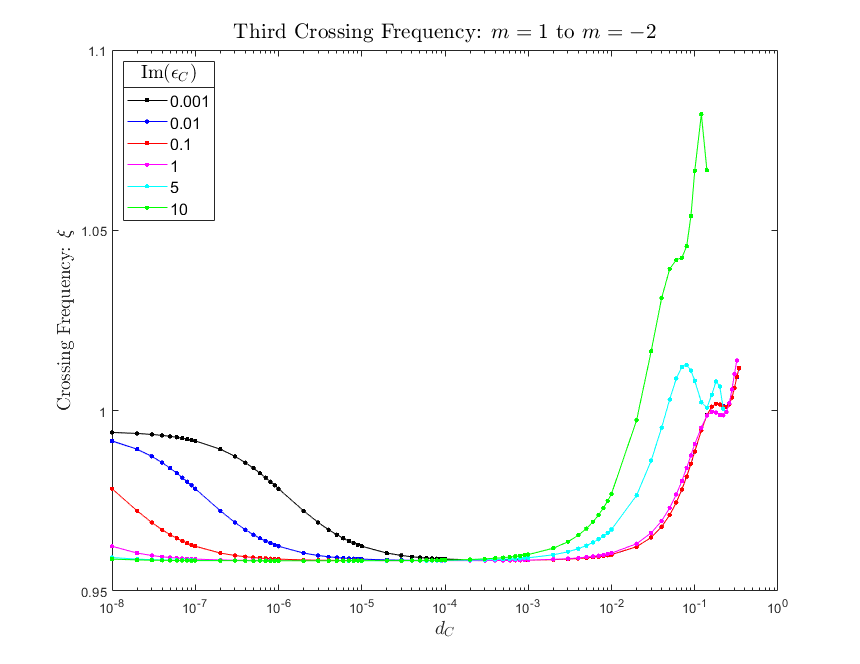}   \label{3_Third_sing_dC} }
	\caption{Behavior of (a) $1^{st}$  (b)$2^{nd}$  (c) $3^{rd}$ crossing point as $d_C$ changes}
	\label{3_dC_Freq}
\end{figure}

We consider a single inversion symmetric quaternary PC made of three different dielectrics, as in Fig.~\ref{QPC}. The optical path to unit cell period ratio is $\gamma=\Gamma/\Lambda = 1.5$. Only one unit cell is used so that the only crossing points occur in the band gaps. We will consider three different systems and examine the lowest three crossings in each case.

In the first study, \textrm{Im}($\epsilon_A$) was allowed to change, keeping all other parameters constant. The remaining parameters used are: \textrm{Re}$(\epsilon_A) = 6$, $\mu_A = 1$, $\epsilon_B = 1$, $\mu_B = 1$, $\epsilon_C = 3$, $\mu_C = 1$. The results are displayed in Fig.~\ref{3_ImA_Freq}. In each plot, the behavior is displayed for four different values of $d_C$: \{0, 0.1, 0.2, 0.3\}. For each particular value of $d_C$, the general trend is for the crossing frequency to start at a lower value for no loss. Then, as  \textrm{Im}($\epsilon_A$) increases, the frequency increases up to some saturation level, $\xi_{\infty}$, marked by horizontal dashed lines. This is because as \textrm{Im}($\epsilon_A$) grows without bound, $d_A\rightarrow0$, in accordance with Eq.~\ref{dA}.  In all three figures, as $d_C$ increases, there is general trend not only for the saturation frequency to be lower, but also for the initial starting value to be closer to it. This makes sense because as $d_C$ increases, $d_A$ will go to zero, for the chosen $n_A$, $n_B$, and $n_C$. The smaller $d_A$ is, the less of an impact its loss component will have on the crossing frequnecy. The rate of convergence appears faster in lower band gaps. Both Figs.~\ref{3_Im_Freq_1} and \ref{3_Im_Freq_2} show this as the horizontal axis is linear; however, for Fig.~\ref{3_Im_Freq_3}, plotting is done on a logarithmic scale  to account for the slower convergence, especially when $d_C=0$. Also in Fig.~\ref{3_Im_Freq_3}, note that when $d_C=0$, the rate of increase temporarily slows around \textrm{Im}$(\epsilon_A)=10$ before increasing again.

In the second study, \textrm{Im}($\epsilon_C$) is allowed to change, keeping all other parameters constant. All other parameters are unchanged from the first study and $\epsilon_A$ is now lossless. Results are shown in Fig.~\ref{3_ImC_Freq} for six different values of $d_C$. Crossing frequencies asymptotically approach constants as \textrm{Im}$(\epsilon_C) \rightarrow0$.  Note that for larger $d_C$, the graphs abruptly end. This is due to the behavior of the layer thicknesses of the corresponding unit cell, given by Eqs.~\ref{dA} and ~\ref{dB}. As $d_C$ becomes larger, $d_A$ becomes negative and thus invalid at smaller values of   \textrm{Im}($\epsilon_C$). The behavior in Fig.~\ref{3_First_sing_ImC} shows the frequency values increasing monotonically whereas in Fig.~\ref{3_Second_sing_ImC} and Fig.~\ref{3_Third_sing_ImC} the frequencies are shown to reach a maximum, at least for smaller $d_C$, before sharply declining.

In the third study , $d_C$ is allowed to change. Other parameters are: $\epsilon_A = 6+10^{-8}i$, $\mu_A = 1$, $\epsilon_B = 1$, $\mu_B = 1$, \textrm{Re}$(\epsilon_C) = 3$, $\mu_C = 1$. The tiny imaginary part of $\epsilon_A$ produces noticeable behavior in the crossing frequency values for $d_C \sim 10^{-7}$ and below. If  \textrm{Im}$(\epsilon_A)$ were zero, then all values would asymptotically approach 0.31682 in Fig.~\ref{3_First_sing_dC}, 0.70029 in Fig.~\ref{3_Second_sing_dC}, and 0.95837 in Fig.~\ref{3_Third_sing_dC}; however, non-zero Im($\epsilon_A$) elevates the crossing point values in odd band gaps and depresses them in even gaps as $d_C\rightarrow0$. Common behavior among all three plots are that as \textrm{Im}$(\epsilon_C)$  becomes larger, the crossing frequency terminates at lower $d_C$. This is only significant for \textrm{Im}$(\epsilon_C)\geq1$ and, as in the second study,  is due to Eq.~\ref{dA} eventually becoming negative, which is unphysical. For larger $d_C$, behavior of first, second, and third crossing frequencies is quite different. In  Fig.~\ref{3_First_sing_dC}, the frequency rises relatively rapidly, more so for larger  \textrm{Im}$(\epsilon_C)$, before leveling off as $d_C$ reaches its maximum value. In  Fig.~\ref{3_Second_sing_dC}, the frequency achieves a maximum before sharply falling. The maximum for \textrm{Im}$(\epsilon_C)<1$ is present but much smaller. In  Fig.~\ref{3_Third_sing_dC}, the crossing point frequencies undergo undulatory behavior. 

\begin{figure}[!ht]
	\subfloat[][]{\includegraphics[width=0.5\columnwidth]{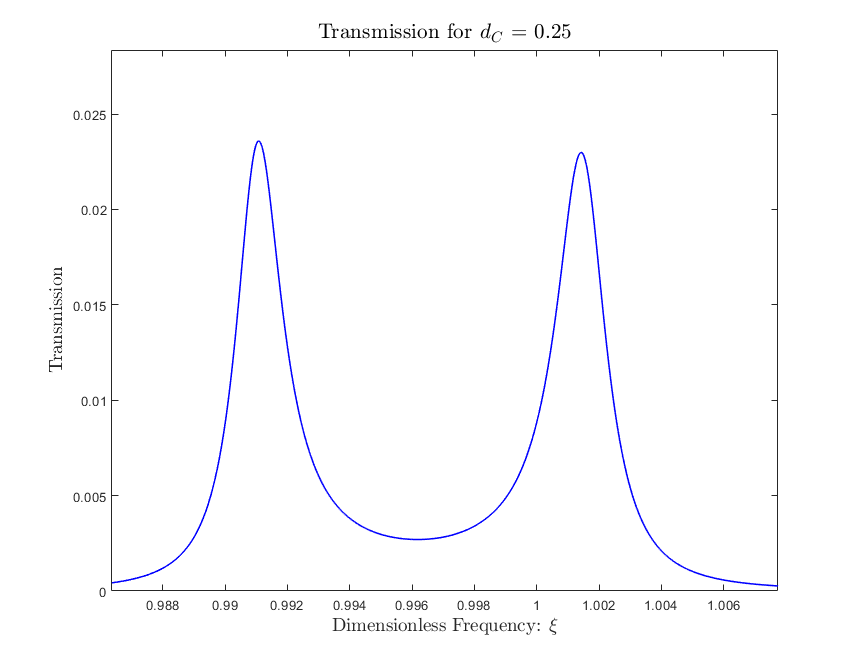}   \label{TwoPeaks} }
	\subfloat[][]{\includegraphics[width=0.5\columnwidth]{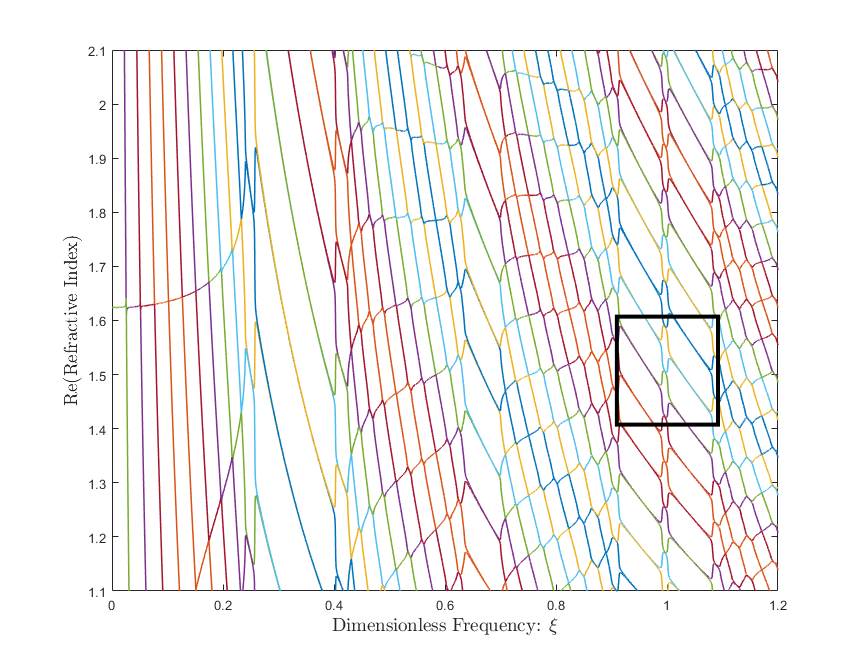}   \label{Eff_bqb1} }\\
	\subfloat[][]{\includegraphics[width=0.5\columnwidth]{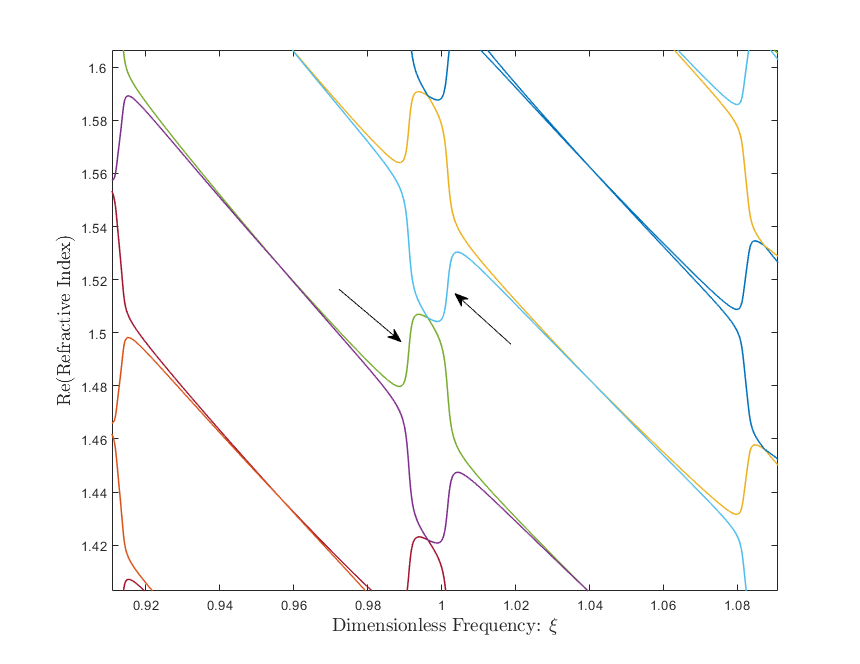}   \label{Eff_bqb2} }	
	\caption{(a) Two transmission peaks in $3^{rd}$  band gap for heterostructure: binary/quaternary/binary. Parameters are: $\gamma = 1.5$, $\epsilon_A = 6 + 0.01i$, $\mu_A = 1$, $\epsilon_B = 1$, $\mu_B = 1$, $\epsilon_C = 3 + 0.01i$, $\mu_C = 1$, $d_C = 0.25$. $d_A$ and $d_B$ are given by Eqs.~\ref{dA} \& \ref{dB}.  (b) Real part of effective refractive index. The boxed region is in (c). (c) The arrows indicate the regions that correspond to the transmission peaks. From left to right, the physcial solution smoothly transitions across the following branches: purple $\rightarrow$ green $\rightarrow$ cyan $\rightarrow$ yellow. }
	\label{Eff_bqb}
\end{figure}

We now consider a heterostructure consisting of a quaternary PC sandwiched between two identical binary PCs (Fig.~\ref{PC_diagram} but with no $C$ layer) \cite {Bianchi2}. All crystals in the structure are inversion symmetric and so is the entire structure. A system with a binary PC surrounding by two quaternary PCs would also be valid; however, a system with an equal number of both cystals would not (\textit{e.g.}quaternary-binary). This is because such system, as a whole is not inversion symmetric depsite the  crystals composing it to being so. We can apply the techniques in Section 2 to this system and find the overall effective refractive index just as for an individual PC. Now though, we do not have the liberty of setting the effective slab thickness equal to a unit cell since now periodicity is broken. We must set the slab thickness equal to the total length of the heterostructure. Since both crystals have the same unit cell period, $L = (2m+1)N \Lambda$, where $m$ is an integer. The effective index profile is similar to that described in Ref.~\cite{Yariv2}, except now spatially  dispersive behavior, caused by the breaking of periodicity via the three different PCs,  is superimposed onto the overall spatial dispersion described previously. In Ref.~\cite{Yariv2}, material dispersion was superimposesd onto spatial dispersion. 

Coupling between the two PC interfaces results in two distinct interface states, displayed in Fig.~\ref{TwoPeaks}, in the $3^{rd}$ band gap for the given parameters: $\gamma = 1.5$, $\epsilon_A = 6 + 0.01i$, $\mu_A = 1$, $\epsilon_B = 1$, $\mu_B = 1$, $\epsilon_C = 3 + 0.01i$, $\mu_C = 1$, $d_C = 0.25$.
If EMT is used to homogenize the entire heterostucture into a uniform slab, the resultant  Re$(n_{eff})$  is displayed in Fig.~\ref{Eff_bqb1}. The boxed region is displayed in Fig.~\ref{Eff_bqb2}. Normally, the band gap is represented with a negative derivative in Re$(n_{eff})$. Note, however, the two regions marked by arrows that show brief instances of normal dispersion. These parts correspond to the two transmssion peaks. Small loss terms were added to some of the permittivities to make the branches smooth across the transition points.  As in the case for an isolated crystal, the band gaps are represented by a large decrease in the index; however, now each branch crossing contains a small resonance, since the heterostructure, as a whole, is not periodic. This break in periodicity also means the exact behavior of the index is now dependent on the number of unit cells in each PC. As mentioned above, this break in periodicity forces us to use the full length of the structure in Eqs.~\ref{n_eff_real} and \ref{n_eff_imag}, leading to a higher number density of branches.  It is still possible to apply to analytical and numerical techniques disccssed in Section 2 to this system, but it may be more tedious.

\section{Conclusion}

We have used effective medium theory to describe some properites of quaternary photonic crystals, both isolated and within heterostructures. For an isolated quaternary PC, the behavior of branch crossing points within the anomalous dispersion region was discussed as various parameters changed. As Im$(\epsilon_A)$ increases, crossing points eventually reach a saturation frequency, that becomes lower for larger $d_C$. In the case of varying Im$(\epsilon_C)$, keeping all else constant, the crossing points are constant for small  Im$(\epsilon_C)$ but display different behavior depending on the point in question. The highest defined crossing point becomes smaller as $d_C$ increases. The roles of Im$(\epsilon_C)$ and $d_C$ are then reversed with $d_C$ now as the independent parameter. It is found that asymptotic behavior as $d_C \rightarrow 0$ is strongly dependent on Im$(\epsilon_A)$. The effective index is then examined for a binary/quaternary/binary heterostructure. It was found that interface states within the band gap are represented as rapid increases in an otherwise decreasing region of the real part of the refractive index.


\clearpage
\begin{thebibliography}{100}

\bibitem{Rytov}
S. M. Rytov,
\emph{Electromagnetic Properties of a Finely Stratified Medium},
Sov. Phys. JETP \textbf{2}, 466 (1956).

\bibitem{Yablonovitch}
E. Yablonovitch,
\emph{Inhibited Spontaneous Emission in Solid-State Physics and Electronics},
Phys. Rev. Lett. \textbf{58}, 2059 (1987).

\bibitem{John}
S. John,
\emph{Strong localization of photons in certain disordered dielectric superlattices},
Phys. Rev. Lett. \textbf{58}, 2486 (1987).

\bibitem{Smith3}
D. R. Smith, W. J. Padilla, D. C. Vier, S. C. Nemat-Nasser, and S. Schultz,
\emph{Composite Medium with Simultaneously Negative Permeability and Permittivity},
Phys. Rev. Lett. \textbf{76}, 4773 (1996)

\bibitem{Lalanne}
P. Lalanne,
\emph{Effective medium theory applied to photonic crystals composed of cubic or square cylinders},
Appl. Opt. \textbf{35}, 5369 (1996)

\bibitem{Kikuta}
H. Kikuta, Y. Ohira, H. Kubo, and K. Iwata,
\emph{Effective medium theory of two-dimensional subwavelength gratings in the non-quasi-static limit},
J. Opt. Soc. Am. A \textbf{15}, 1577 (1998)

\bibitem{Chern}
R. L. Chern, and Y. T. Chen,
\emph{Effective parameters for photonic crystals with large dielectric contrast},
Phys. Rev. B \textbf{80}, 075118 (2009)

\bibitem{Ono}
Y. Ono,
\emph{Transmittance analysis of three-dimensional photonic crystals by the effective medium theory},
Appl. Opt. \textbf{45}, 131 (2006)

\bibitem{Datta}
S. Datta, C. T. Chan, K. M. Ho, and C. M. Soukoulis,
\emph{Effective dielectric constant of periodic composite structures},
Phys. Rev. B \textbf{48}, 14936 (1993)

\bibitem{Tang}
S. Tang, B. Zhu, M. Jia, Q. He, S. Sun, Y. Mei, L. Zhou,
\emph{Effective-medium theory for one-dimensional gratings},
Phys. Rev. B \textbf{91}, 174201 (2015)

\bibitem{Kameda}
S. Kameda, A. Mizutani, and H. Kikuta,
\emph{Effective Medium Theory for Calculating Reflectance from Metal-Dielectric Multilayered Structure},
Jpn. J. Appl. Phys. \textbf{51}, 042202 (2012)

\bibitem{Yariv2}
A. Yariv, and P. Yeh,
\emph{Electromagnetic propagation in periodic stratified media. II. Birefringence, phase matching, and x-rays lasers},
J. Opt. Soc. Am. \textbf{67}, 438 (1977)

\bibitem{Yong}
G. Ji-Yong, C. Hong, L. Hong-Qiang, and Z. Ye-Wen,
\emph{Effective permittivity and permeability of one dimensional dielectric photonic crystal within a band gap},
Chinese Phys. \textbf{17}, 2544 (2008)

\bibitem{Liu1}
Y. Liu, S. Guenneau, and B. Gralak,
\emph{Causality and passivity properties of effective parameters of electromagnetic multilayered structures},
Phys. Rev. B \textbf{88}, 165104 (2013)

\bibitem{Liu2}
Y. Liu, S. Guenneau, and B. Gralak,
\emph{Artificial dispersion via high-order homogenization: magnetoelectric coupling and magnetism from dielectric layers},
Proc. R. Soc. London. Ser. A \textbf{469}, 20130240 (2013)

\bibitem{Boedecker}
G. Boedecker, and C. Henkel,
\emph{All-frequency effective medium theory of a photonic crystal},
Opt. Express \textbf{11}, 1590 (2003)

\bibitem{Zhao}
Q. Zhao, J. Liu, D. Gao, K. You, X. Wang, H. M. Leung, T. K. Yung, R. Zhang, and W. Y. Tam,
\emph{Reflection phase of photonic bands in finite bi-directional 1D photonic crystals using an effective medium approach},
OSA Continuum \textbf{1}, 332 (2018)

\bibitem{Gao}
J. Liu, D. Gao, W. Mao, Q. Zhao, H. Ma, Y. Wang, X. Wang, T. K. Yung, and W. Y. Tam,
\emph{Characterization of free-standing 1D photonic crystals using an effective medium approach},
Opt. Lett. \textbf{44}, 4853 (2019)

\bibitem{Pei}
T. H. Pei, and Y. T. Huang,
\emph{The equivalent structure and some optical properties of the periodic-defect photonic crystal},
J. Appl. Phys. \textbf{109}, 073104 (2011)

\bibitem{Bianchi}
N. Bianchi, and L. Kahn,
\emph{Topological Photonic States at a 1-D Binary-Quaternary Interface},
(to be published) 	arXiv:1910.02920 

\bibitem{Soukoulis}
P. Markoš, and C. M. Soukoulis,
\emph{Wave Propagation From Electrons to Photonic Crystals and Left-Handed Materials},
(Princeton Unversity Press, 2008).

\bibitem{Yariv1}
A. Yariv, and P. Yeh,
\emph{Optical Waves in Crystals: Propagation and Control of Laser Radiation},
(John Wiley \& Sons, 1984). 

\bibitem{Smith1}
D. R. Smith, S. Schultz, P. Markos, and C. M. Soukoulis,
\emph{Determination of effective permittivity and permeability of metamaterials from reflection and transmission coefficents},
Phys. Rev. B \textbf{65}, 195104 (2002)

\bibitem{Smith2}
D. R. Smith, D. C. Vier, Th. Koschny, and C. M. Soukoulis,
\emph{Electromagnetic parameters retrieval from inhomogeneous metamaterials},
Phys. Rev. E \textbf{71}, 036617 (2005)

\bibitem{Chen2}
X. Chen, T. M Grzegorczyk, B. I. Wu, J. Pacheco Jr, and J. A. Kong,
\emph{Robust method to retrieve the constitutive effective parameters of metamaterials},
Phys. Rev. E \textbf{70}, 016608 (2004)

\bibitem{Notomi}
M. Notomi,
\emph{Theory of light propagation in strongly modulated photonic crystals: Refractionlike behavior in the vicinity of the photonic band gap},
Phys. Rev. B \textbf{62}, 10696 (2000)

\bibitem{Dowling}
J. P. Dowling, and C. M. Bowden,
\emph{Anomalous index of refraction in photonic bandgap materials},
J. Mod. Opt. \textbf{41}, 345 (1994)

\bibitem{Jeong}
D. Y. Jeong, Y. H. Ye, and Q. M. Zhang,
\emph{Effective optical properties associated with wave propagation in photonic crystals of finite length along the propagation direction},
J. Appl. Phys. \textbf{92}, 4194 (2002)

\bibitem{Bianchi2}
N. Bianchi, and L. Kahn,
\emph{Optical States in a 1-D Superlattice with Multiple Photonic Crystal Interfaces},
(to be published)   	arXiv:2001.01165

\end{thebibliography}
\end{document}